\documentclass[aps,prl,twocolumn]{revtex4}
\usepackage{graphicx}
\usepackage{textcomp}
\newcommand{\comment}[1]{}

\begin{document}

\title{Critical Behavior of a Trapped Interacting Bose Gas \footnote{This is the author's version of the work. It is posted here by permission of the AAAS for personal use, not for redistribution. The definitive version was published in SCIENCE, 315, (16 March 2007), doi:10.1126/science.1138807.}}
\author {T. Donner$^1$, S. Ritter$^1$, T. Bourdel$^1$, A. {\"O}ttl$^1$, M. K{\"o}hl$^{1,2\dag}$, T. Esslinger$^1$
\\
\normalsize{$^1$Institute of Quantum Electronics, ETH Z\"{u}rich,
CH--8093 Z\"{u}rich, Switzerland}\\
\normalsize{$^2$Cavendish Laboratory, University of Cambridge, JJ Thomson Avenue,}\\
\normalsize{Cambridge CB3 0HE, United Kingdom}\\
\normalsize{$^\dag$To whom correspondence should be addressed;
E-mail:  koehl@phys.ethz.ch.}  }

\begin{abstract}
The phase transition of Bose-Einstein condensation is studied in
the critical regime, when fluctuations extend far beyond the
length scale of thermal de Broglie waves. Using matter-wave interference
we measure the correlation length of these critical fluctuations
as a function of temperature. The diverging behavior of the correlation length above the critical
temperature is observed, from which we determine the critical
exponent of the correlation length for a trapped, weakly
interacting Bose gas to be $\nu=0.67\pm 0.13$. This measurement has direct implications
for the understanding of second order phase transitions.
\end{abstract}

\maketitle

Phase transitions are among the most dramatic phenomena in nature, when
minute variations in the conditions controlling a system can trigger a
fundamental change of its properties. For example, lowering the
temperature below a critical value creates a finite magnetization
of ferromagnetic materials or, similarly, allows for the
generation of superfluid currents. Generally, a transition takes
place between a disordered phase and a phase exhibiting
off-diagonal long-range order which is the magnetization or the
superfluid density in the above cases. Near a second-order phase
transition point the fluctuations of the order parameter are so
dominant that they completely govern the behavior of the system on
all length scales \cite{zinnjustin1996}. In fact, the large scale
fluctuations in the vicinity of a transition already indicate the
onset of the phase on the other side of the transition.

Near a second-order phase transition, macroscopic quantities show a
universal scaling behavior which is characterized by critical
exponents \cite{zinnjustin1996} that depend only on
general properties of the system, such as its dimensionality,
symmetry of the order parameter or the range of interaction.
Accordingly, phase transitions are
classified in terms of universality classes. Bose-Einstein
condensation in three dimensions, for example, is in the same
universality class as a three-dimensional XY magnet. Moreover, the
physics of quantum phase transitions occurring at zero temperature
can often be mapped on thermally driven phase transitions in
higher spatial dimensions.

The phase transition scenario of Bose-Einstein condensation in a
weakly interacting atomic gas is unique as it is free of
impurities and the two-body interactions are precisely known. As
the gas condenses, trapped bosonic atoms of a macroscopic number
accumulate in a single quantum state and can be described by the
condensate wave function, the order parameter of the
transition. However, it has proven to be experimentally difficult
to access the physics of the phase transition itself. In
particular, the critical regime has escaped observation as it
requires an extremely close and controlled approach to the
critical temperature. Meanwhile, advanced theoretical methods have
increased the understanding of the critical regime in a gas of
weakly interacting bosons
\cite{baym1999,arnold2001a,kashurnikov2001,arnold2001}. Yet, the
theoretical description of the experimental situation, a Bose gas
in a harmonic trap, has remained unclear.

\begin{figure}[htbp]
  \includegraphics[width=1\columnwidth]{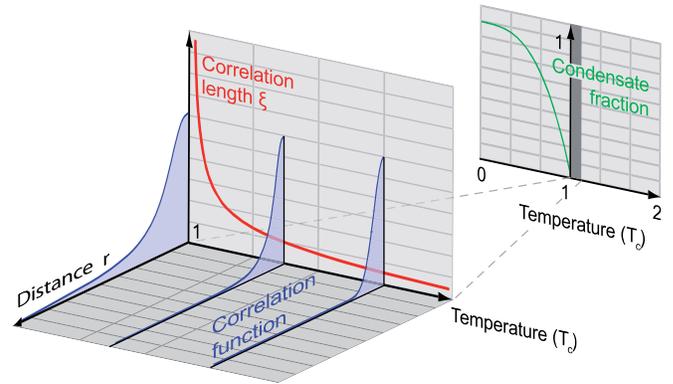}
  \caption{
  Schematics of the correlation function and the correlation length close to the phase transition temperature of Bose-Einstein condensation.
  Above the critical temperature $T_c$ the condensate fraction is zero (see right) and for $T\gg T_c$
  the correlation function decays approximately as a Gaussian on
  a length scale set by the thermal de Broglie wavelength $\lambda_{dB}$. As the temperature approaches the critical temperature
  long-range fluctuations start to govern the system and the correlation length $\xi$ increases dramatically.
  Exactly at the critical temperature $\xi$ diverges and the correlation function decays algebraically for $r>\lambda_{dB}$ (Eq \ref{corr_fit}).}
  \label{fig1}
\end{figure}

We report on a measurement of the correlation length of a
trapped Bose gas within the critical regime just above the
transition temperature. The visibility of a matter-wave interference pattern gives
us direct access to the first order correlation function. Exploiting our experimental temperature
resolution of 0.3\,nK (corresponding to $2 \times 10^{-3}$
of the critical temperature) we observe the divergence of the
correlation length and determine its critical exponent $\nu$. This
direct measurement of $\nu$ through the single particle density
matrix complements the measurements of other critical exponents in
liquid Helium \cite{goldner1993,adriaans1994,lipa2003}, which is
believed to be in the same universality class as the weakly
interacting Bose gas.

In a Bose gas the physics of fluctuations of the order parameter
is governed by different length scales. Far above the phase
transition temperature, classical thermal fluctuations dominate.
Their characteristic length scale is determined by the thermal de
Broglie wavelength $\lambda_{dB}$ and the correlation function can
be approximated by $\langle \Psi^\dagger(r)\Psi(0)\rangle \propto
\exp(-\pi r^2/\lambda_{dB}^2)$ with $r$ being the separation of
the two probed locations \cite{naraschewski1999} (Fig 1). Nontrivial fluctuations of the order parameter $\Psi$
close to the critical temperature become visible when their length
scale becomes larger than the thermal de Broglie wavelength. The
density matrix of a homogeneous Bose gas for $r>\lambda_{dB}$ can
be expressed by the correlation function \cite{huang1987,foot0}
\begin{equation}
\langle \Psi^\dagger(r)\Psi(0)\rangle \propto\frac{1}{r}
\exp(-r/\xi), \label{corr_fit}
\end{equation}
where $\xi$ denotes the correlation length of the order parameter.
The correlation length $\xi$ is a function of temperature $T$ and
diverges as the system approaches the phase transition (Fig
1). This results in the algebraic decay of the
correlation function with distance $\langle
\Psi^\dagger(r)\Psi(0)\rangle \propto 1/r$ at the phase
transition. The theory of critical phenomena predicts a divergence
of $\xi$ according to a power-law
\begin{equation}
\xi\propto |(T-T_c)/T_c|^{-\nu},
\end{equation}
where $\nu$ is the critical exponent of the correlation length and
$T_c$ the critical temperature. The value of the critical exponent
depends only on the universality class of the system.

While for non-interacting systems the critical exponents can be
calculated exactly \cite{zinnjustin1996,foot1}, the presence of
interactions adds richness to the physics of the system. Determining the value of
the critical exponent using Landau's theory of phase transitions
results in a value of $\nu=1/2$ for the homogeneous system. This
value is the result of both a classical theory and a mean field
approximation to quantum systems. However, calculations initially
by Onsager \cite{onsager1944} and later the techniques of the
renormalization group method \cite{zinnjustin1996} showed that
mean field theory fails to describe the physics at a phase
transition. Very close to the critical temperature -- in the
critical regime -- the fluctuations become strongly correlated and
a perturbative or mean-field treatment becomes impossible making
this regime very challenging.

We consider a weakly interacting Bose gas with density $n$ and the
interaction strength parameterized by the s-wave scattering length
$a=5.3$\,nm in the dilute limit $n^{1/3}a\ll 1$. In the critical regime
mean-field theory fails because the fluctuations of $\Psi$ become
more dominant than its mean value. This can be determined by the
Ginzburg criterion $\xi
> \lambda_{dB}^2/(\sqrt{128}\pi^2 a)\simeq 0.4\,\mu$m \cite{giorgini1996,foot2}.
Similarly, these enhanced fluctuations are responsible for a
nontrivial shift of the critical temperature of Bose-Einstein
condensation
\cite{baym1999,arnold2001a,kashurnikov2001,prokofev2004}. The critical regime of
a weakly interacting Bose gas offers an intriguing possibility to
study physics beyond the usual mean-field approximation
\cite{niu2006} which has been observed in cold atomic gases only
in reduced dimensionality before
\cite{stoferle2004,paredes2004,kinoshita2004,hadzibabic2006}.

In our experiment, we let two atomic beams, which
originate from two different locations spaced by a distance $r$
inside the trapped atom cloud, interfere. From the visibility of the
interference pattern the first order correlation function \cite{bloch2000}
of the Bose gas above the critical temperature and the correlation length $\xi$
can be determined.
\begin{figure}[htbp]
  \includegraphics[width=1\columnwidth]{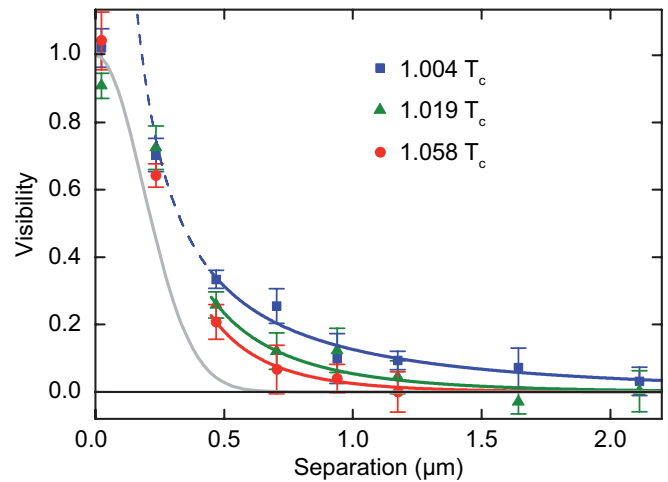}
  \caption{  Spatial correlation function of a trapped Bose gas close to the
  critical temperature. Shown is the visibility of a matter wave
  interference pattern originating from two regions separated by
  $r$ in an atomic cloud just above the transition
  temperature. The gray line is a Gaussian with a width given
  by the thermal de Broglie wavelength $\lambda_{dB}$ which
  changes only marginally for the temperature range considered here. The
  experimental data show phase correlations extending far beyond the scale set by
  $\lambda_{dB}$. The solid line is a fit proportional to $\frac{1}{r}e^{-r/\xi}$ for $r>\lambda_{dB}$.
  Each data point is the mean of on average 12 measurements, the error-bars are $\pm$\,SD.}
  \label{fig2}
\end{figure}

We prepare a sample of $4 \times 10^{6}$ $^{87}$Rb atoms in the $|F = 1, m_F = -1
\rangle$ hyperfine ground state in a magnetic trap\cite{ottl2006}. The trapping
frequencies are ($\omega_{x}$, $\omega_{y}$, $\omega_{z}$) = 2$\pi \times (39, 7, 29) \mathrm{Hz}$, where $z$
denotes the vertical axis. Evaporatively cooled to just below the critical temperature the sample reaches a
density of $n=2.3 \times 10^{13}$\,$\mathrm{cm}^{-3}$
giving an elastic collision rate of $90\,\mathrm{s^{-1}}$. The temperature is
controlled by holding the atoms in the trap for a defined period of time during which energy is transferred
to the atoms due to resonant stray light, fluctuations of the
trap potential or background gas collisions. From absorption images we determine
the heating rate to be $4.4\pm0.8\,\mathrm{nK/s}$. Using this
technique we cover a range of temperatures from $0.001\,
<(T-T_c)/T_c < 0.07$ over a time scale of seconds.

For output coupling of the atoms we use microwave frequency fields to spin-flip the atoms into the magnetically untrapped state $|F = 2, m_F = 0 \rangle$. The resonance condition for this transition is given by the local magnetic field and the released atoms propagate downwards due to gravity. The regions of output coupling are chosen symmetrically with respect to the center of
the trapped cloud and can be approximated by horizontal planes
spaced by a distance $r$ \cite{bloch2000}. The two released atomic
beams interfere with each other. For the measurement we typically
extract $4\times10^4$ atoms over a time scale of 0.5\,s which is
approximately $1\%$ of the trapped sample. We detect the
interference pattern in time with single atom resolution using a
high finesse optical cavity, placed 36\,mm below the center of the
magnetic trap. An atom entering the cavity mode decreases the
transmission of a probe beam resonant with the cavity. Due to the
geometry of our apparatus, only atoms with a transverse momentum
$(p_x, p_y) \simeq 0$ are detected which results in an overall
detection efficiency of $1\%$ for every atom output coupled from
the cloud. From the arrival times of the atoms we find the
visibility $\mathcal{V}(r)$ of the interference pattern
\cite{bourdel2006}. From repeated measurements with different
pairs of microwave frequencies we measure $\mathcal{V}(r)$ with
$r$ ranging from  0 to 4 $\lambda_{dB}$ where $\lambda_{dB}\simeq
0.5\,\mu$m.

With the given heat rate a segmentation of the acquired visibility data into time bins of
$\Delta t =72$\,ms length allows for a temperature resolution of
0.3\,nK which corresponds to 0.002\,$T_c$. The time bin length was
chosen to optimize between shot-noise limited determination of the
visibility from the finite number of atom arrivals and
sufficiently good temperature resolution. For the analysis we have
chosen time bins overlapping by $50\%$.

Figure 2 shows the measured visibility as a function of
slit separation $r$ very close to the critical temperature $T_c$.
We observe that the visibility decays on a much longer length
scale than predicted by the thermal de Broglie wavelength
$\lambda_{dB}$. We fit the long distance tail $r>\lambda_{dB}$
with Eq 1 (solid line) and determine the correlation
length $\xi$. The strong temperature dependence of the correlation
function is directly visible. As $T$ approaches $T_c$ the
visibility curves become more long ranged and similarly the
correlation length $\xi$ increases. The observation of long-range
correlations shows how the size of the correlated regions strongly
increases as the temperature is varied only minimally in the
vicinity of the phase transition.

\begin{figure}[htbp]
  \includegraphics[width=1\columnwidth]{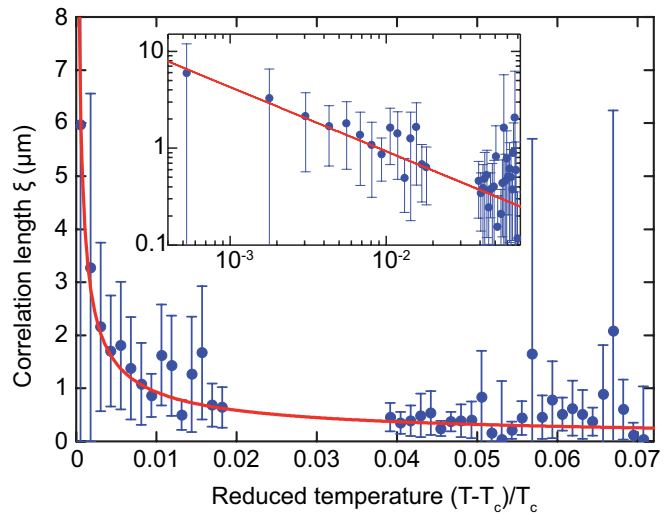}
  \caption{  Divergence of the correlation length $\xi$ as a function of temperature.
  The red line is a fit of Eq 2 to the data with $\nu$ and $T_c$ as free
  parameters. Plotted is one data set for a specific temporal offset $t_0$.
  The error bars are $\pm$\,SD, according to fits to Eq 1. They also reflect the scattering between different data sets.
  Inset: Double logarithmic plot of the same data.}
  \label{fig3}
\end{figure}

Figure 3 shows how the measured correlation length $\xi$
diverges as the system approaches the critical temperature.
Generally, an algebraic divergence of the correlation length is
predicted. We fit our data with the power law according to
Eq 2, leaving the value of $T_c$ as a free fit parameter, which has a typical relative error of $5 \times 10^{-4}$.
Therefore our analysis is independent of an exact calibration of
both temperature and heating rate provided that the heating rate
is constant. The resulting value for the critical exponent is
$\nu=0.67\pm0.13$. The value of the critical exponent is averaged
over 30 temporal offsets $0<t_0<\Delta t$ of the analyzing
time bin window and the error is the reduced $\chi^2$ error.
Systematic errors on the value of $\nu$ could be introduced by the
detector response function. We find the visibility for a pure
Bose-Einstein condensate to be $100\%$ with a statistical error of
$2\%$ over the range of $r$ investigated. This uncertainty of the
visibility would amount to a systematic error of the critical
exponent of 0.01 and is neglected as compared to the statistical
error. The weak singularity of the heat capacity near the $\lambda$-transition \cite{zinnjustin1996}
results in an error of $\nu$ of less than 0.01.

Finite size effects are expected when the correlation length is large
\cite{damle1996a,lipa2000} and they may lead to a slight underestimation of
$\nu$ for our conditions. Moreover, the harmonic confining
potential introduces a spatially varying density. The phase
transition takes place at the center of the trap and
non-perturbative fluctuations are thus expected within a finite
radius $R$ \cite{arnold2001}. Using the Ginzburg criterion as
given in \cite{giorgini1996} we find $R \approx 10\,\mathrm{\mu
m}$, whereas the rms size of the thermal cloud is 58\,$\mu$m. The longest
distance we probe in our experiment is $2\,\mathrm{\mu m}$ which
is well below this radius $R$.

So far, in interacting systems the critical exponent $\nu$ has
been determined for the homogeneous system. The
$\lambda$-transition in liquid Helium is among
the most accurately investigated systems at criticality. One
expects to observe the same critical exponents
despite the fact that the density differs by ten orders of
magnitude. In the measurements with liquid Helium the critical
exponent of the specific heat $\alpha$ has been measured
in a spaceborne experiment \cite{lipa2003}. Through the
scaling relation $\alpha=2-3\nu$ the value of the critical
exponent $\nu \simeq 0.67$ is inferred being in
agreement with theoretical predictions
\cite{campostrini2001,burovski2006}. Alternatively, the exponent
$\zeta\simeq 0.67$ (which is related to the superfluid density
$\rho_s=|\Psi|^2$ instead of the order parameter $\Psi$) can be
measured directly in second sound experiments in liquid Helium
\cite{goldner1993,adriaans1994}. Due to an argument by Josephson
\cite{josephson1966} it is believed that $\nu=\zeta$, however, a
measurement of $\nu$ directly through the density matrix has so
far been impossible with Helium.

The long-range behaviour of the
correlation function of a trapped Bose gas in the critical regime reveals the behavior of a phase
transition in a weakly interacting system. Our measured value for the
critical exponent does not coincide with that obtained by a simple
mean-field model or with that of an ideal non-interacting Bose
gas. The value is in good agreement with the expectations of
renormalization group theory applied to a homogeneous gas of
bosons and with measurements using strongly interacting superfluid
Helium.

\nocite{acknoledgements}

\bibliographystyle{Science}

\begin{thebibliography}{10}
\bibitem{foot0}
In the general d-dimensional case $\langle \Psi^\dagger(r)\Psi(0)\rangle \propto
r^{-(d-2+\eta)} \exp(-r/\xi)$ with $\eta$ being a critical
exponent. The calculated and measured values of $\eta$ are on the
order of $10^{-2}$ and therefore we neglect $\eta$ in our
analysis.
\bibitem{foot1}
For a noninteracting gas the critical exponent can be calculated to be $\nu=1$
for homogeneous systems and $\nu=1/2$ for harmonically trapped systems.
\end{thebibliography}

\end{document}